    \documentstyle[twocolumn,epsbox]{jpsj}

\def\mbf(#1){\mbox{\boldmath $#1$}}
\def\runtitle{Chiral Superconducting State
}
\def\runauthor{Kazuhiro {\sc Kuboki} }

\title
{Effect of Band Structure on the Symmetry of Superconducting States
}

\author{Kazuhiro {\sc Kuboki}\footnote{E-mail:
kuboki@phys.sci.kobe-u.ac.jp}}

\inst{Department of Physics, Kobe University, Kobe 657-8501, Japan}

\recdate{ \hskip3.0truecm }

\abst{
Effects of the band structure on the symmetry of superconducting (SC) 
states are studied.  
For a square lattice system with a nearest-neighbor attractive interaction, 
SC states with various symmetries are found by changing the band structure, 
or, the shape of the Fermi surface. 
The spin-triplet ($(p_x + ip_y)$-wave) and spin-singlet ($d$- or $s$-wave) 
SC states, and states with their coexistence ($d + ip_y$, $s + ip_y$) 
can be stabilized within the same type of interaction. 
The stability of interlayer-pairing states with line nodes is also examined, 
and its relation to the SC state of Sr$_2$RuO$_4$ is discussed.
}

\kword{ $p$-wave superconductivity,  band structure, Ginzburg-Landau theory, 
Sr$_2$RuO$_4$
}

\begin{document}
\sloppy
\maketitle


The superconducting (SC) state of Sr$_2$RuO$_4$ attracts much attention, 
since it is likely to have a spin-triplet pairing 
symmetry.\cite{Maeno,Rice,Ishida,Luke,Mack} 
The triplet Cooper pairs in $^3$He are formed due to the ferromagnetic 
spin fluctuation, so that it may be natural to assume that a 
SC state in Sr$_2$RuO$_4$ is also realized by the same mechanism. 
However, recent neutron scattering experiments showed that 
the ferromagnetic (${\vec q} = 0$) spin fluctuation is not large, 
but the peak is located near 
${\vec q} = (\pm 2\pi/3, \pm 2\pi/3)$,\cite{neutron} 
consistent with  band structure calculations.\cite{Mazin} 
In view of this fact it is proposed that the antiferromagnetic spin fluctuation 
may lead to the spin-triplet SC state.\cite{Sato, Ogata}
Similar results have also been obtained in different 
contexts.\cite{Kagan,Chubukov} 

In this article we study the effect of the band structure on the symmetry of 
SC states. 
First we treat a single-band tight-binding model on a square lattice 
using a mean-field approximation (MFA). 
We find that 
the spin-triplet and the spin-singlet SC states, together with states with 
their coexistence can occur for the same type of interaction simply by 
changing the shape of the Fermi surface.\cite{tsuchi}
Micnas {\it et al}.\cite{Micnas} studied the stability of various SC 
states in a model 
similar to ours, but they have determined only the bare $T_{\rm c}$. 
Namely they solved only the linearized self-consistency equations. 
We will solve the self-consistency equations without linearization 
to determine the phase diagram, and clarify the reason for the 
change of the symmetry of the SC state as the band structure is changed. 

Experimental results on Sr$_2$RuO$_4$ seem to indicate that the SC state in
this system  has a line (or lines) of 
nodes.\cite{Nishi,Ishida2,Bonalde,Matsuda,Lupien} 
In order to explain these results theoretically,\cite{Miyake,Graf,Maki,Kuroki} 
Hasegawa {\it et al}.\cite{Hase} proposed 
an interlayer-pairing state which has horizontal lines of nodes 
based on a symmetry argument. 
We will examine the stability of this type of 
interlayer-pairing state when the in-plane and the interlayer interactions 
compete.

First we consider a
tight-binding model on a square lattice 
whose Hamiltonian is given by 
\begin{equation}\begin{array}{rl}
H  = & \displaystyle -\sum_{i,j,\sigma} t_{ij} 
c^\dagger_{i\sigma}c_{j\sigma} 
- \mu \sum_{i\sigma} c^\dagger_{i\sigma} c_{i\sigma} \\ 
& \\
- & \displaystyle  \sum_{i,j} V_{ij} 
n_{i\uparrow}n_{j\downarrow}
\end{array}\end{equation}
where $\mu$ is the chemical potential, and $t_{ij}$ is defined as 
\begin{equation}
t_{ij} = \displaystyle 
t \sum_{\delta=\pm {\hat x},\pm {\hat y}} \delta_{i,j+\delta}
+ t' \sum_{\delta=\pm {\hat x} \pm {\hat y}} \delta_{i,j+\delta}.
\end{equation} 
Namely,  $t$ ($t'$) is the transfer integral for the 
(next) nearest-neighbor sites and ${\hat x}$ (${\hat y}$) is the unit 
vector in the $x$ ($y$) direction (lattice constant is taken to be unity). 
Similarly the nearest-neighbor attractive interaction  $V_{ij}$ is defined as 
$V_{ij} = V \sum_{\delta=\pm {\hat x}, \pm {\hat y}}
\delta_{i,j+\delta}$ ($V > 0$). 
This Hamiltonian is decoupled by a standard mean-field procedure
\begin{equation}\begin{array}{rl}
n_{i\uparrow}n_{j\downarrow} = & 
c^\dagger_{i\uparrow}c_{i\uparrow} 
c^\dagger_{j\downarrow}c_{j\downarrow} \\
& \\
\to & \displaystyle
\Delta_{ij}  c^\dagger_{j\downarrow}c^\dagger_{i\uparrow} 
+ \Delta_{ij}^{*} c_{i\uparrow}c_{j\downarrow} 
- |\Delta_{ij}|^2
\end{array}\end{equation} 
with $\Delta_{ij} \equiv \langle c_{i\uparrow}c_{j\downarrow} \rangle$ 
being the SC order parameter (OP). 
On the square lattice 
the $d_{x^2-y^2}$- ($\Delta_d$), extended $s$- ($\Delta_s$), 
$p_x$- ($\Delta_{p_x}$) and $p_y$-wave  ($\Delta_{p_y}$) symmetries 
are possible for the nearest-neighbor interaction, 
and the corresponding OP's are defined as
\begin{equation}\begin{array}{rl}
\Delta_d(i) = & \displaystyle  (\Delta_{i,i+x} + \Delta _{i,i-x} 
- \Delta_{i,i+y} - \Delta_{i,i-y})/4  \\
\Delta_s(i) = & \displaystyle  (\Delta_{i,i+x} + \Delta_{i,i-x} 
+ \Delta_{i,i+y} + \Delta_{i,i-y})/4 \\
\Delta_{p_{x(y)}}(i)  = & \displaystyle 
i (\Delta_{i,i+x(y)} - \Delta_{i,i-x(y)})/2 . 
\end{array}\end{equation} 
Assuming that these OP's are uniform (i.e., independent of $i$) we obtain 
the following self-consistency equations: 
\begin{equation}\begin{array}{rl}
\Delta_{d(s)} = & \displaystyle \frac{V}{4N}\sum_k 
\omega_{d(s)}(k)\frac{\Delta_k}{E_k}\tanh \Bigl(\frac{E_k}{2T}\Bigr) \\
& \\
\Delta_{p_{x(y)}} = &  \displaystyle \frac{V}{2N}\sum_k 
\omega_{p_{x(y)}}(k)\frac{\Delta_k}{E_k}\tanh \Bigl(\frac{E_k}{2T}\Bigr)
\end{array}\end{equation} 
where $N$ and $T$ are the total number of lattice sites and the temperature, 
respectively,  and 
\begin{equation}\begin{array}{rl}
E_k = & \displaystyle \sqrt{\xi_k^2 + |\Delta_k|^2}  \\
& \\
\xi_k = & \displaystyle  -2t(\cos k_x + \cos k_y) -4t'\cos k_x \cos k_y -\mu \\
& \\
\Delta_k = & \displaystyle 2 \sum_{j=d,s,p_x,p_y} \omega_j(k) \Delta_j
\end{array}\end{equation} 
with 
\begin{equation}\begin{array}{rl}
\omega_d(k) = & \displaystyle \cos k_x - \cos k_y, \\ 
\omega_s(k) = & \displaystyle \cos k_x + \cos k_y, \\
\omega_{p_{x(y)}} = & \displaystyle \sin k_{x(y)}.
\end{array}\end{equation} 
In the following we will solve the self-consistency equations to 
determine the phase giagram in the plane of $T$ and $\mu$. 
These equations are solved by the iteration method, starting from 
various sets of initial values. When several solutions are obtained 
for the same set of parameters ($\mu$, $T$),  
the state with the lowest free energy is adopted as the true one. 
Here we note that the SC long-range order cannot exist at finite temperature 
in a purely two dimensional (2D) system.
However, $T_c$ obtained within the MFA in purely 2D systems can give 
a reasonable estimate of $T_c$ in the presence of small three dimensionality.

In Fig.1 we show the phase diagram in the plane of $T$ and $\mu$ 
(or, the electron density), for $t =1$,  $t' = 0$ and $V = 1.5$.
(Due to the particle-hole symmetry, the result for $-\mu$ is the same as that for 
$\mu$, and so it is not shown.)
It is seen that a $d_{x^2-y^2}$-wave SC state is stabilized near 
half-filling ($\mu \sim 0$), 
while an extended $s$-wave state occurs at high (and low) densities 
($\mu \sim \pm 4t$). In the region between $d$- and $s$-wave states 
spin-triplet $(p_x \pm ip_y)$-wave states appear. 
The $(p_x + ip_y)$- and the  $(p_x - ip_y)$-states are degenerate 
but different states, and they transform each other 
under parity ${\cal P}$ and time-reversal ${\cal T}$ transformation. 
Then the system breaks ${\cal P}$ and ${\cal T}$ symmetries spontaneously, and 
these states are usually denoted as the chiral $p$-wave SC states. 
Near the boundary between triplet $(p_x \pm ip_y)$ and singlet ($d$ or $s$) 
states we find states where the spin-singlet and  the spin-triplet 
OP's coexist. 
These coexisting states,  
$(d \pm ip_y)$-  and $(s \pm ip_y)$-states, are degenerate with 
$(d \pm ip_x)$-  and $(s \pm ip_x)$- states, respectively.\cite{mixture}
The $(d \pm ip_x \pm ip_y)$ and $(s \pm ip_x \pm ip_y)$ 
states are slightly higher in energy so that they are only local minima 
of the free energy. 
There is no reason (regarding symmetry) which precludes the coexistence 
of spin-triplet and spin-singlet SCOP's, and it is the energy that 
decides which state should appear. 
Actually the coexistence of $d$- and $p$-wave OP's  has been 
found in superconductor/(anti)ferromagnet bilayer systems,  
where the proximity effect induces the imbalance of spin-up and spin-down 
electron densities.\cite{prox}

The above results show that both triplet and singlet SC states can occur 
with the same type of interaction. 
In order to understand this we use the Ginzburg-Landau (GL) theory 
by expanding the free energy with respect to $\Delta$'s\cite{SigUe}
\begin{equation} 
\begin{array}{rl}
{\cal F}_\Delta   = & \displaystyle \frac{1}{S} \int d^2r \Bigl(\sum_{j=d,s,p_x,p_y} 
\big[\alpha_j|\Delta_j|^2 + \beta |\Delta_j|^4 
\big] \\ 
& \\ 
+ & \displaystyle \sum_{i \not= j} \big[\gamma_{ij}^{(1)} |\Delta_i|^2 
|\Delta_j|^2 + \gamma_{ij}^{(2)} \big(\Delta_i^2\Delta_j^{*2} + c.c.
\big)\big]\Bigr) 
\end{array}
\end{equation}
where the gradient and higher order terms are discarded, and $S$ is the 
area of the system.  
As $T$ is decreased the OP with the highest (bare) transition temperature 
$T_c^{(0)}$ ($\alpha(T_c^{(0)}) =  0$) appears. The explicit forms of 
$\alpha$'s are given as 
\begin{equation}\begin{array}{rl}
\alpha_{d(s)} = & \displaystyle 4V \Bigl(1 - \frac{V}{N}\sum_k 
\omega_{d(s)}^2(k)\frac{\tanh(\xi_k/2T)}{2\xi_k}\Bigr) \\ 
& \\
\alpha_{p_{x(y)}} = & \displaystyle 2V \Bigl(1 - \frac{V}{N}\sum_k
\omega_{p_{x(y)}}^2(k) \frac{\tanh(\xi_k/2T)}{\xi_k}\Bigr).
\end{array}\end{equation}
The Fermi surface (FS) near the band edge ($\mu \sim \pm 4t$) 
is close to the $\Gamma$ point or ${\mib k} = (\pm \pi, \pm \pi)$, 
so that $|\omega_s(k)|$ is large on the FS and thus $\Delta_s$ is favored. 
On the other hand the FS at half-filling is the square 
connecting four points $(\pm \pi, 0), (0,\pm \pi)$, 
and $\omega_s(k)$ vanishes there. Then $\Delta_s$ is suppressed near 
half-filling and $\Delta_d$ is favored.
For intermediate $\mu$ ($\mu \sim \pm 2t$), 
the FS comes close to the points $k_x =\pm \pi/2$ 
or $k_y = \pm \pi/2$ so that $|\omega_{p_{x(y)}}|$ can be large.
Then the $p$-wave states have the highest $T_c$ in the region between 
$d$- and $s$-wave states
($p_x$ and $p_y$ states are degenerate). 

When more than one $\alpha$ become negative there may be a coexistence of 
several OP's. In this case $\gamma$ terms will play important roles. 
We can explicitly show that 
$\beta_i > 0$, $\gamma_{ij}^{(1)} > 0$ and $\gamma_{ij}^{(2)} > 0$ 
($i,j = d, s, p_x, p_y$).  
The fact $\gamma_{ij}^{(2)} > 0$ indicates that the OP's 
would form complex rather than real combinations (if they coexist), 
and in this case the nodes are removed and the system gains more 
condensation energy. 
This is the reason why the chiral ($p_x \pm ip_y$)-state (rather than 
($p_x \pm p_y$)-state) appears. 
The coexisting states also have the complex combinations of OP's 
due the same reason. 
In Fig.1 the region of $(d + ip_y)$-state is much wider than that of 
$(s + ip_y)$-state. This can be understood as follows. 
The nodes in the $d$-wave state can be removed by the introduction 
of the $ip_x$ component, while in the $s$-state there is already a full gap 
so that the lowering of the energy due to the second OP is much smaller. 

Whether or not the above argument is correct can be tested by considering 
the case of $t = 0$, $t' \not =1$, i.e., with only the next-nearest-neighbor 
hopping terms. 
The Fermi surface at half-filling consists of the lines 
$k_x = \pm \pi/2$ and $k_y = \pm \pi/2$, so that the 
$(p_x \pm ip_y)$-state should be most favored near half-filling. 
This is actually the case as shown in Fig.2. 
Here the chiral SC state appears at and near half-filling, and $d$- or 
extended $s$-wave state occurs away from half-filling.

Next we examine the stability of interlayer-pairing SC states 
in a tetragonal system with a weak interlayer interaction. 
The Hamiltonian in this case is 
\begin{equation}\begin{array}{rl}
H =  & \displaystyle H_{2D} + H_\perp \\
& \\
H_\perp = & \displaystyle 
- t_\perp \sum_i \sum_{\delta=\pm {\hat z}} c^\dagger_i c_{i+\delta}
- V_\perp \sum_i \sum_\delta n_{i\uparrow} n_{i+\delta,\downarrow}
\end{array}\end{equation}
where the summation on $\delta$ in the second term is taken over 
$\delta=\pm {\hat z} \pm {\hat x}, \pm {\hat z} \pm {\hat y}$ 
and $H$ in eq.(1) is redefined as $H_{2D}$. 
Here ${\hat z}$ denotes the unit vector in the $z$-direction with a lattice 
constant $c$.  
For the $t_\perp$-term we simply take the nearest-neighbor hopping, but 
we do not consider the nearest-neighbor interaction because of the 
following reason. 
The interaction terms with $\delta = \pm {\hat z}$ (with coupling constant 
$V_\perp^{(0)}$) may lead to the OP of the form $\Delta_{i,i \pm z}$. 
This OP is invariant under the rotation around the $z$-axis so that it is
decoupled from the in-plane OP (denoted as $\Delta_\parallel$) if 
$\Delta_\parallel$ has a $d$- or $p$-wave symmetry. 
Then $V_\perp^{(0)}$ of the order of $V$ (in-plane 
coupling constant) 
is necessary to stabilize $\Delta_{i,i \pm z}$. 
When $\Delta_\parallel$ has an $s$-wave symmetry, it couples to 
$\Delta_{i,i\pm z}$ and then the latter becomes finite even for an infinitesimal 
$V_\perp^{(0)}$. 
Since we are not interested in the $s$-wave case, 
we do not consider $V_\perp^{(0)}$ in the following.

We decouple $H_\perp$ using the same procedure as in 2D case. 
The possible symmetries of the OP's are 
\begin{equation}\begin{array}{rl}
(\alpha) & \sin k_x \cos k_zc, \ \  \sin k_y \cos k_zc  \\
(\beta) & \cos k_x \sin k_zc, \ \  \cos k_y \sin k_zc  \\
(\gamma) & \sin k_x \sin k_zc,  \ \  \sin k_y \sin k_zc   \\
(\delta) & \cos k_x \cos k_zc, \ \  \cos k_y \cos k_zc.   
\end{array}\end{equation}
Here ($\alpha$) and ($\beta$) (($\gamma$) and ($\delta$)) are 
spin-triplet (spin-singlet) states. 
(Note the states with $x$ and $y$ interchanged are degenerate.)
Among these eight OP's we are interested in the states in ($\alpha$), 
since their complex combinations, i.e., 
($\sin k_x \pm i\sin k_y)\cos k_zc$ are proposed to describe the SC state of 
Sr$_2$RuO$_4$.\cite{Hase} 
This state has horizontal lines of nodes, and this behavior is 
consistent with the experimental results. 
For $t_\perp = 0$ all interlayer pairing OP's ($\Delta_\perp$) 
do not couple to $\Delta_\parallel$,
while for $t_\perp \not= 0$ some of $\Delta_\perp$ may have a bilinear 
coupling (in the sense of GL theory)  to $\Delta_\parallel$ if both OP's  
have the same symmetry.  In the latter case 
$\Delta_\perp$ can be finite once $V_\perp \not= 0$. 
In view of this we take $t=1, t'=0, V_1=1.5$ and $\mu=-2$, since   
$\Delta_\parallel$ has the $(p_x \pm ip_y)$-symmetry for these values of 
parameters, and OP's of $(\alpha)$ may couple to $\Delta_\parallel$.
In the following we consider $(\sin k_x + i\sin k_y)\cos k_zc$-  and 
$(\sin k_x + i\sin k_y)\sin k_zc$-wave OP's, and denote them 
as $\Delta_\perp^{(c)}$ and $\Delta_\perp^{(s)}$, respectively. 
Then we calculate $\Delta_\parallel$, $\Delta_\perp^{(c)}$ and 
$\Delta_\perp^{(s)}$ self-consistently as functions of $V_\perp$.

In Fig.3 the results for $t_\perp = 0$ at $T = 0$ is shown. 
In this case $\Delta_\parallel$ and $\Delta_\perp$ are not coupled. 
Since the FS has no warping along the $z$-axis,  
it is not energetically favorable to introduce  
the second component of OP ($\Delta_\perp$) because all parts of the 
Fermi surface are already gapped by $\Delta_\parallel$. 
Thus $\Delta_\perp$ is absent for $V_\perp < V_{\rm c}$ 
($V_{\rm c} \sim V$). 
When $V_\perp > V_{\rm c}$, $\Delta_\perp^{(c)}$ and $\Delta_\perp^{(s)}$ 
appear simultaneously, since these two states are degenerate when 
$t_\perp$. 
Then the state for large $V_\perp$ has 
($\Delta_\perp^{(c)} + i\Delta_\perp^{(s)}$)-symmetry and is fully gapped. 
The in-plane OP is excluded due to the same reason which suppresses 
$\Delta_\perp$ when $V_\perp < V_{\rm c}$.
Thus we conclude that the state does not have nodes irrespective of the value 
of  $V_\perp$ if $t_\perp = 0$.

Next we consider the case $t_\perp \not= 0$. 
In this case $\Delta_\parallel$ couples to $\Delta_\perp^{(c)}$. 
Then the latter can be finite once  $V_\perp$ becomes finite. 
Now the gap function is  
\begin{equation}
\Delta_k = 
(\sin k_x + i\sin k_y)(\Delta_\parallel + \Delta_\perp^{(c)}\cos k_zc 
+ i\Delta_\perp^{(s)}\sin k_zc). 
\end{equation}
Since $\Delta_\perp^{(c)}$ is induced by the bilinear coupling to 
$\Delta_\parallel$, their relative phase is 
either $0$ or $\pi$,  while $\Delta_\perp^{(s)}$ favors a phase $\pm \pi/2$ relative 
to them due to $\gamma$ terms. 
In  Fig.4 the results are shown for $t_\perp = 0.4$. 
Here $|\Delta_\parallel| > |\Delta_\perp| $ (and $\Delta_\perp^{(s)} = 0$)
for small $V_\perp$ so that there is a full gap. 
For large values of $V_\perp$, 
$|\Delta_\perp^{(c)}| > |\Delta_\parallel|$. 
However,  $i\Delta_\perp^{(s)}$ component 
appears before $|\Delta_\perp^{(c)}|$ exceeds $|\Delta_\parallel|$.
Then the state is again fully gapped except an accidental case where 
$|\Delta_\parallel^{(c)}| = |\Delta_\perp|$. 
We have also examined other values of $t_\perp$, and 
the SC state (with $(p_x \pm ip_y)$-symmetry 
for the in-plane OP) always has a full gap except an accidental case, 
unless $t_\perp$ becomes comparable to $t$.

In summary we have studied the symmetry of the SC states 
in a single-band tight-binding 
model with an attractive interaction between nearest-neighbor sites.  
It is shown that the spin-triplet and 
the spin-singlet SC states, and even their coexistence can occur as the 
band structure is changed. 
These results can be understood by considering the change of the shape of 
the Fermi surface. 
The present result implies that the band structure is an important factor 
to determine the symmetry of the SC state. 
We have also examined the stability of interlayer-pairing states with 
line nodes. 
These states are difficult to be stabilized in a model with such a simple 
band structure (Fermi surface) as that used in the present work. 
Experimental results of Sr$_2$RuO$_4$ seem to indicate that the SC state 
has a spin-triplet symmetry, 
and that there is a line (or lines)  of nodes in the excitation gap. 
In order to give consistent interpretations of these facts, 
it would be necessary to consider the model which takes into account the 
more realistic crystal (and band) structure of Sr$_2$RuO$_4$.\cite{Hase2}
This problem will be examined elsewhere. 

\smallskip
				
The author is grateful to M. Sigrist, Y. Hasegawa, Y. Tanaka and  H. Fukuyama 
for useful discussions. 
He also thanks Ken Yokoyama for letting him know Ref.10, 11. 
This work was supported in part by a Grant-in-Aid for Scientific 
Research from the Ministry of Education, Science, Sports and Culture 
of Japan.



\bigskip
\noindent 
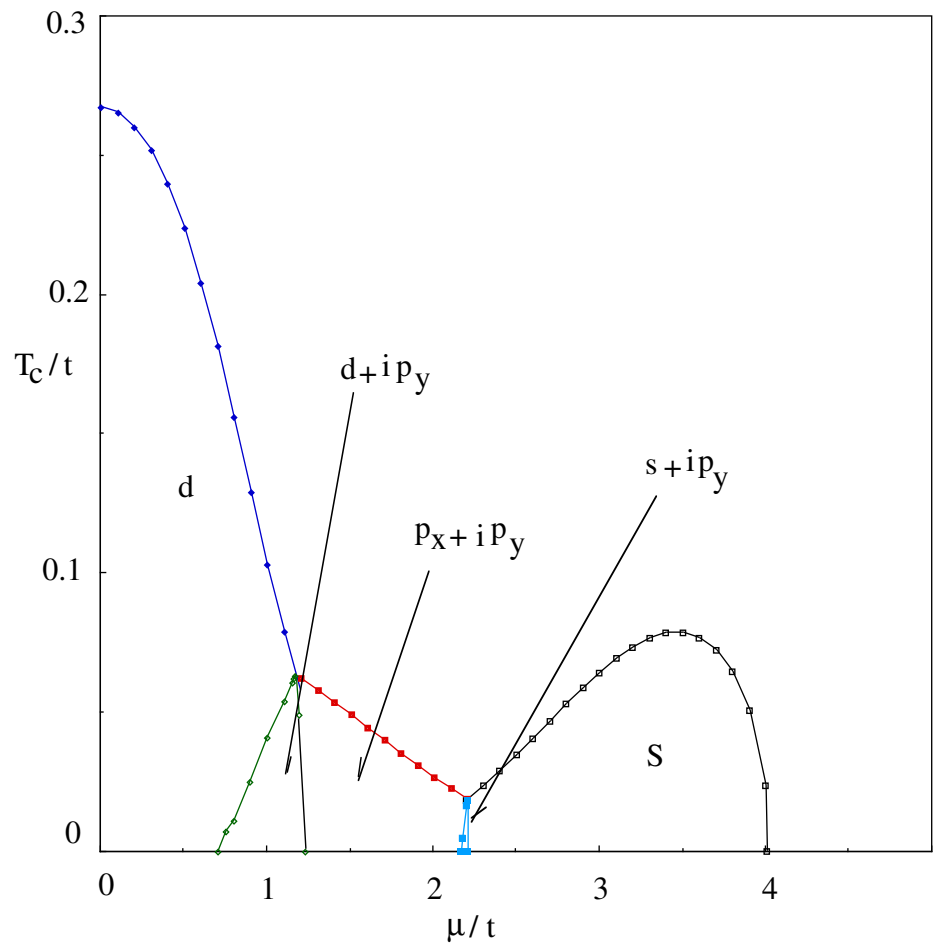
  {\bf Fig. 1}  
  Phase diagram in the plane of $T$ and $\mu$.
  Parameters used are $t = 1, t' =0$ and $V = 1.5$. 

\bigskip
\noindent 
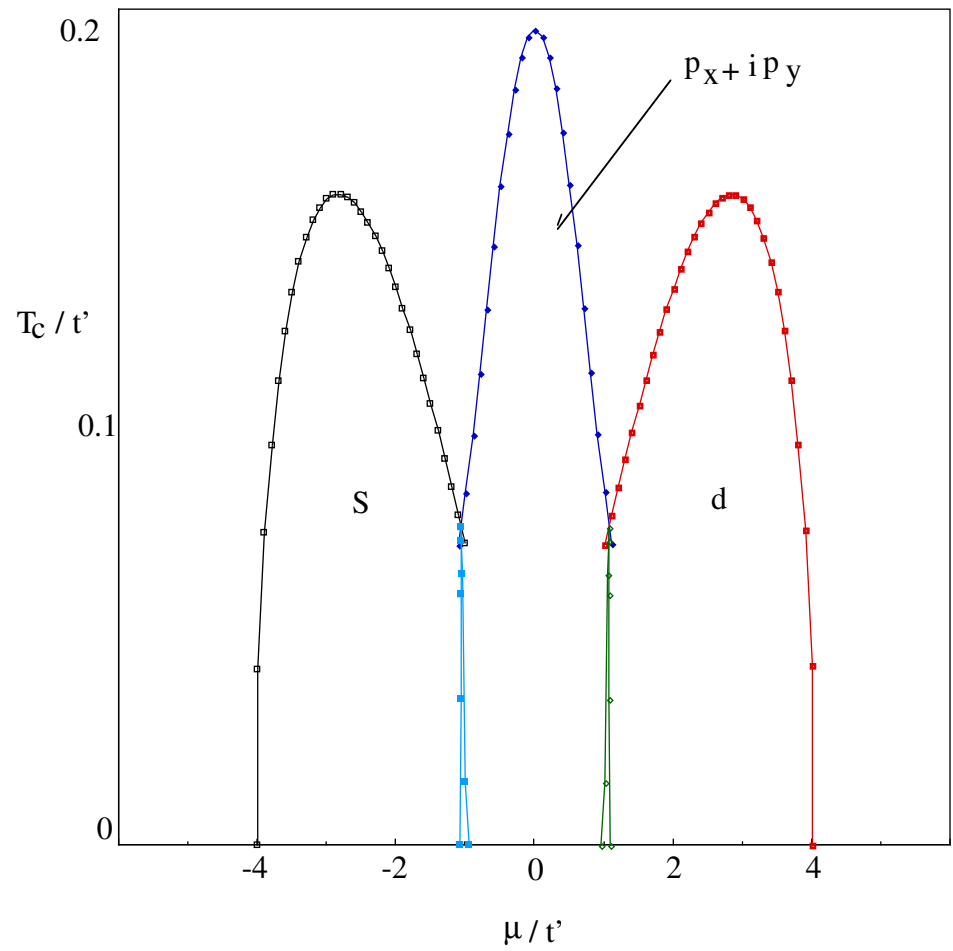
  {\bf Fig. 2}   
   Phase diagram in the plane of $T$ and $\mu$.
   Parameters used are $t = 0, t' =1$ and $V = 1.5$. 
   A narrow region between $s$ ($d$) and $p_x +ip_y$ states is an $s + ip_y$ 
  ($d + ip_y$) state.  
  
\bigskip
\noindent 
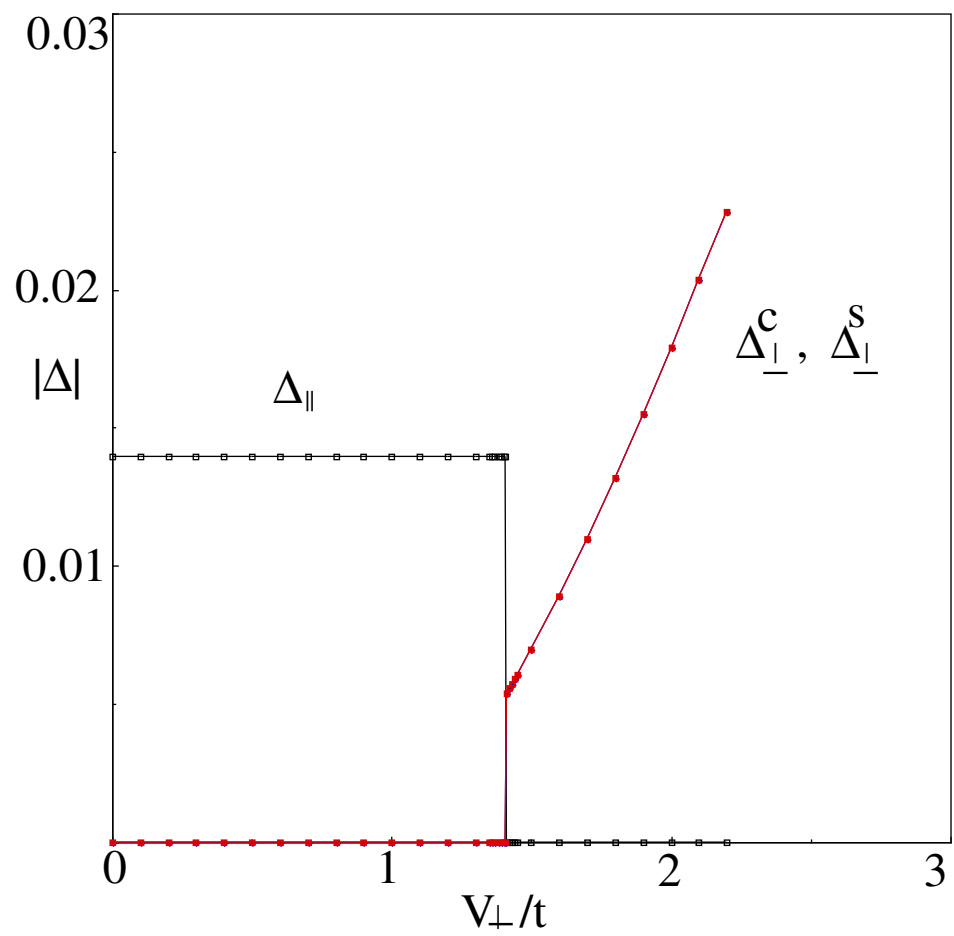
  {\bf Fig. 3}    
  The $V_\perp$ dependence of SC order 
  parameters for  
  $t=1, t'=0, V=1.5, \mu=-2, T = 0$ and $t_\perp=0$. 
  Note that all order parameters are non-dimensional. 
  
\bigskip
\noindent 
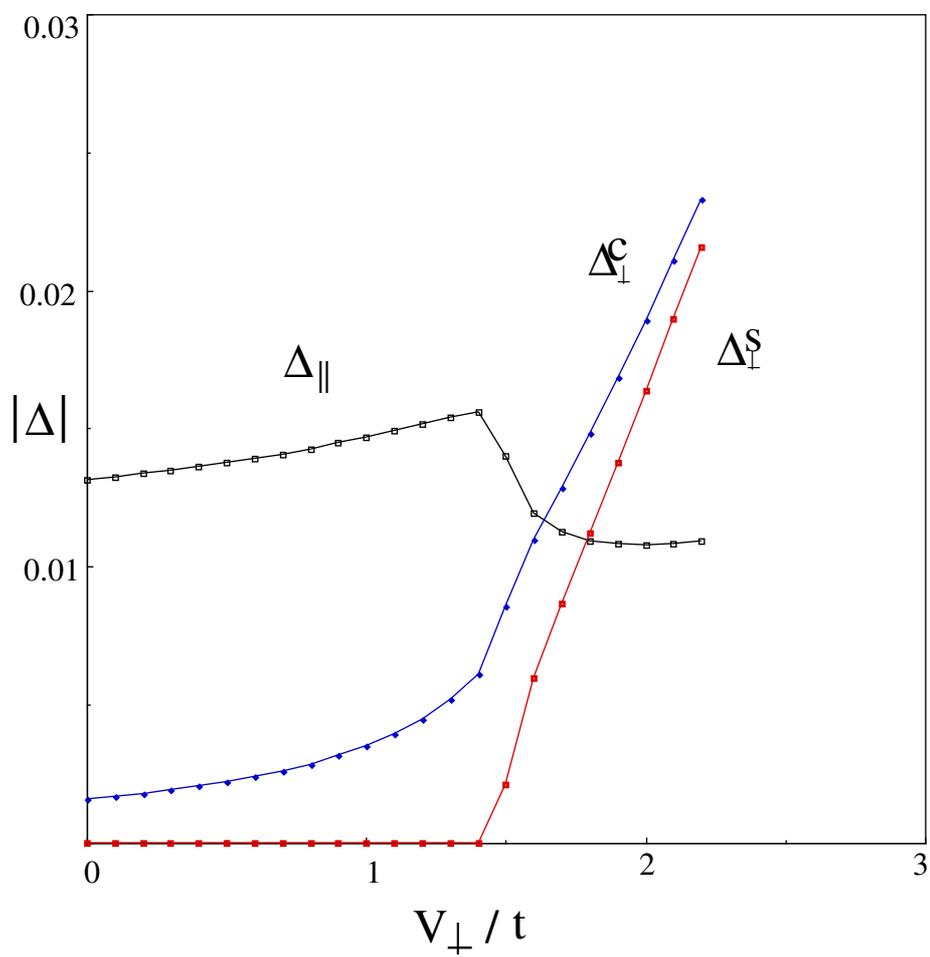
  {\bf Fig. 4}    
  The $V_\perp$ dependence of SC order 
  parameters for $t=1, t'=0, V=1.5, \mu=-2, T = 0$ and $t_\perp=0.4$. 
  Note that all order parameters are non-dimensional.


\begin{thebibliography}{99}

\bibitem{Maeno} Y. Maeno, H. Hashimoto, K. Yoshida, S. Nishizaki, T. Fujita, 
J. G. Bednorz and F. Lichtenberg: Nature {\bf 372} (1994) 532. 

\bibitem{Rice} T. M. Rice and M. Sigrist: J. Phys. Condens. Matter {\bf 7} 
(1995) L643. 

\bibitem{Ishida} K. Ishida, H. Mukuda, Y. Kitaoka, K. Asayama, Z. Q. Mao, 
Y. Mori and Y. Maeno: Nature {\bf 396} (1998) 658. 

\bibitem{Luke} G. M. Luke, Y. Fudamoto, K. M. Kojima, M. I. Larkin, 
J. Merrin, B. Nachumi, Y. J. Uemura, Y. Maeno, Z. Q. Mao, H, Nakamura 
and M. Sigrist: Nature {\bf 394} (1998) 558. 

\bibitem{Mack} A. P. Mackenzie, R. K. W. Haselwimmer, A. w. Tyler, 
G. G. Lonzarich, Y. Mori, S. Nishizaki and Y. Maeno: Phys. Rev. Lett. 
{\bf 80} (1998) 161. 

\bibitem{neutron} Y. Sidis, M. Braden, P. Bourges, B. Hennion, S. Nishizaki, 
Y. Maeno and Y. Mori: Phys. Rev. Lett. {\bf 83} (1999) 3320. 

\bibitem{Mazin} I. I. Mazin and D. J. Singh: Phys. Rev. Lett. {\bf 82} 
(1999) 4324. 


\bibitem{Sato} M. Sato and M. Kohmoto: J. Phys. Soc. Jpn. {\bf 69} (2000) 3505. 

\bibitem{Ogata}T. Kuwabara and M. Ogata:  Phys. Rev. Lett. {\bf 85} (2000) 4586. 

\bibitem{Kagan} M. Yu. Kagan and T. M. Rice: J. Phys. Condens. Matter 
{\bf 6} (1994) 3771. 

\bibitem{Chubukov} A. V. Chubukov: Phys. Rev. B{\bf 48} (1993) 1097. 

\bibitem{tsuchi} 
The SC state in a similar model  
within the spin-singlet symmtery was studied in: 
H. Tsuchiura, Y. Tanaka and Y. Ushijima: J. Phys. Soc. Jpn. 
{\bf 64} (1995) 922. 


\bibitem{Micnas} R. Micnas, J. Ranninger and S. Robaszkiewicz: 
Rev. Mod. Phys. {\bf 62} (1990) 113.

\bibitem{Nishi}  S. Nishizaki, Y. Maeno and Z.Q. Mao: J. Phys. Soc. Jpn. 
{\bf 69} (2000) 572. 

\bibitem{Ishida2} K. Ishida, H. Mukuda, Y. Kitaoka, Z.Q. Mao, Y. Mori and 
Y. Maeno: Phys. Rev. Lett. {\bf 84} (2000) 5387. 

\bibitem{Bonalde} I. Bonalde, B.D. Yanoff, M.B. Salamon, D.J. van Harlingen, 
E.M.E. Chia, Z.Q. Mao and Y. Maeno:  Phys. Rev. Lett. {\bf 85} (2000) 4775. 

\bibitem{Matsuda} K. Izawa, H. Takahashi, H. Yamaguchi, Y. Matsuda, M. Suzuki, 
T. Sasaki, T. Fukase, Y. Yoshida, R. Settai and Y. Onuki:  cond-mat/0012137.

\bibitem{Lupien} C. Lupien, W.A. MacFarlane, C. Proust, L. Taillefer, 
Z.Q. Mao and Y. Maeno:  cond-mat/0101319.

\bibitem{Miyake} 
K. Miyake and O. Narikiyo: Phys. Rev. Lett. {\bf 83} (1999) 1423.

\bibitem{Graf} M.J. Graf and A.V. Balatsky: Phys. Rev. B {\bf 62} (2000) 9697. 

\bibitem{Maki} T. Dahm, H. Won and K. Maki: cond-mat/0006301.

\bibitem{Kuroki} K. Kuroki, M. Ogata, R. Arita and H. Aoki: cond-mat/0101077. 

\bibitem{Hase} Y. Hasegawa, K. Machida and M. Ozaki: 
J. Phys. Soc. Jpn {\bf 69} (2000) 336.

\bibitem{mixture} A tiny amount of $s$-wave  ($d$-wave) component is mixed 
in the $d + ip_y$  ($s + ip_y$) state. This is because the $x$- and 
the $y$- directions are not equivalent in this state, 
and so the absolute values of OP's for both directions 
are not the same. 
  
  
\bibitem{prox}
K. Kuboki: J. Phys. Soc. Jpn. {\bf 68} (1999) 3150. 

\bibitem{SigUe} For a review see M. Sigrist and K. Ueda, 
Rev. Mod. Phys., {\bf 63}, 239 (1991).
 
\bibitem{Hase2} Y. Hasegawa; private communications. 
  


\end{thebibliography}
\end{document}